\documentclass[aps,prl,floats,twocolumn,superscriptaddress,amssymb,amsmath,showpacs]{revtex4}
\usepackage{graphicx}

\begin{document}
\title{Nonuniversal finite-size scaling in anisotropic systems}

\author{X.S. Chen}
\affiliation{Institute of Theoretical Physics, Chinese Academy of
Sciences, P.O. Box 2735, Beijing 100080, China}
\affiliation{Institute of Theoretical Physics, Aachen University,
D-52056 Aachen, Germany}
\author{V. Dohm}
\affiliation{Institute of Theoretical Physics, Aachen University,
D-52056 Aachen, Germany}

\date{October 25, 2004}

\begin{abstract}

We study the bulk and finite-size critical behavior of the O$(n)$
symmetric $\varphi^4$ theory with spatially anisotropic
interactions of non-cubic symmetry in $d<4$ dimensions. In such
systems of a given $(d,n)$ universality class, two-scale factor
universality is absent in bulk correlation functions, and
finite-size scaling functions including the Privman-Fisher scaling
form of the free energy, the Binder cumulant ratio and the Casimir
amplitude are shown to be nonuniversal. In particular it is shown
that, for anisotropic confined systems, isotropy cannot be
restored by an anisotropic scale transformation.

\end{abstract}
\maketitle

A basic tenet in the physics of critical phenomena is the notion
of a universality class. It is characterized by the dimensionality
$d$ of the system and by the number $n$ of the components of the
order parameter. (See, e.g., the review article \cite{privman-1}.)
Within a certain $(d,n)$ universality class, the universal
quantities (critical exponents, amplitude ratios and scaling
functions) are independent of microscopic details, such as the
particular type of (finite-range or van der Waals type)
interactions or the lattice structure \cite{statement}. This
implies that a given universality class includes both spatially
isotropic and anisotropic systems.

Once the universal quantities of a universality class are known
the asymptotic critical behavior of very different systems (e.g.,
fluids and magnets) is believed to be known completely provided
that only {\it two} nonuniversal amplitudes $A_1$ and $A_2$ are
given. This property is known as {\it two-scale factor
universality} or {\it hyperuniversality} \cite{privman-1,stauffer,
aharony-etc}. In terms of the singular part of the reduced bulk
free energy density $F_s / V k_B T \equiv f_s (t, h)$ above $T_c$,
\begin{equation}
\label{gleichungbulk} f_s (t,h) = A_1 t^{d \nu} W (A_2 h t ^{-
\beta \delta})
\end{equation}
with $W (0) = 1$ and $t = (T - T_c) / T_c \ll 1$, this property
can be stated as \cite{privman-fisher}
\begin{equation}
\label{gleichung0} \lim_{t \to 0 +} f_s (t, 0) \xi^d = Q (d, n) =
{\rm universal} \;.
\end{equation}
Thus the amplitude $\xi_0=(Q / A_1)^{1/d}$ of the correlation
length $\xi=\xi_0  t^{- \nu}$ at zero ordering field $h$ is not an
independent amplitude but is universally related to $A_1$. The
validity of two-scale factor universality has been established by
the renormalization-group (RG) theory on the basis of an {\it
isotropic} Hamiltonian with short-range interactions below the
upper critical dimension $d^* = 4$ \cite{aharony-etc} but no
general proof has been given for the anisotropic case.

In this paper we study the critical behavior of systems with a
spatial anisotropy of non-cubic symmetry within a given $(d,n)$
universality class. An example is an Ising ferromagnet with an
isotropic nearest-neighbor (NN) coupling $J > 0$ and an
anisotropic next-nearest-neighbor (ANNN) coupling $J'$ on a
simple-cubic lattice. In some range of $J'/J$ this model has the
same type of critical behavior as the ordinary $(J' = 0)$ Ising
model. We shall show that for such systems Eq. (\ref{gleichung0})
must be generalized to
\begin{equation}
\label{gleichung3a} \lim_{t \to 0 +} f_s (t, 0) \prod_{i = 1}^d
\xi^{(i)} = Q (d, n) = {\rm universal}
\end{equation}
where $\xi^{(i)}=\xi_0^{(i)} t^{- \nu}$ are the correlation
lengths associated with the principal directions of the
anisotropic system and where $Q (d,n)$ is the same universal
quantity for both isotropic and anisotropic systems. (For $d = 2,
n = 1$, this is already known for the Ising model with anisotropic
NN (ANN) interactions $J_x \neq J_y$ \cite{stauffer}.) There are,
in general, $d+1$ nonuniversal bulk amplitudes $\xi_0^{(1)},
\ldots, \xi_0^{(d)}, A_2$ whose ratios are also nonuniversal. Note
that there still exists a unique critical exponent $\nu (d, n)$
that is identical for isotropic and anisotropic systems within the
same $(d,n)$ universality class \cite{griffiths, bruce,
wu,campostrini,NNN}.

A different type of critical behavior exists in the so-called {\it
strongly} anisotropic systems
\cite{binder-etc,hucht,henkel,caracciolo} where not only {\it
amplitudes} depend on the spatial directions but also the critical
exponents (e.g., $\nu_\parallel$ and $\nu_\perp$) depend on the
direction. These systems do not belong to the $(d,n)$ universality
class of ordinary critical points and our analysis will not
include such types of anisotropy.

While Eq. (\ref{gleichung3a}) is a natural generalization of Eq.
(\ref{gleichung0}) we shall call attention to the intriguing
problem of {\it finite-size} effects in anisotropic systems. For
simplicity we shall confine ourselves to the case of periodic
boundary conditions in rectangular $L_1 \times L_2 \times ...
\times L_d$ block geometries (including $L^d$ cubic geometry and
$\infty^{d-1} \times L$ film geometry). There have been several
studies of this problem in the past
\cite{hucht,nightingale,privman-fisher,kim,hu, yurishchev}. We
shall only briefly comment on the more complicated case of
anisotropic confined systems with non-periodic boundary conditions
\cite{hu,cardy-1983,barber,indekeu,karevski,aharony}.

It has been hypothesized \cite{privman-fisher} that two-scale
factor universality holds not only for {\it bulk} systems but also
for {\it confined} systems, except that the finite-size scaling
functions depend on the geometry and on the boundary conditions.
For example, for a system in a cube of volume $L^d$ with periodic
boundary conditions, the singular part of the reduced free energy
density $f_s (t, h, L)$ near bulk $T_c$ was predicted to have the
asymptotic scaling form for large $L$ \cite{privman-fisher}
\begin{equation}
\label{gleichung1} f_s (t, h, L) = L^{-d} Y_{cube} (C_1 t L^{1 /
\nu}, C_2 h L^{\beta \delta/\nu})
\end{equation}
where the function $Y_{cube} (x, y)$ is universal and where $C_1$
and $C_2$ are the only nonuniversal parameters. A similar ansatz
was made for the correlation length $\xi_\parallel$ in a $L^{d-1}
\times \infty$ cylinder \cite{privman-fisher}. As a consequence,
the amplitude $Y_{cube}(0,0)$ and the Binder cumulant ratio
\cite{privman-1,binder-1981,privman-1990}
\begin{equation}
\label{gleichung5}U=\frac{1}{3} \left[\left(\partial^4 Y_{cube} /
\partial y^4 \right) / \left(\partial^2 Y_{cube} /
\partial y^2 \right)^2 \right]_{y = 0, x = 0}
\end{equation}
are predicted to be universal. (For example, they should be
independent of the ratio $J'/J$.) The scaling form
(\ref{gleichung1}), if extended to realistic geometries and
boundary conditions, has far-reaching consequences for measurable
quantities \cite{privman-1,dohm-1}. In particular the prediction
of a universal character of the Casimir amplitude
\begin{equation}
\label{gleichung6} \Delta = (d - 1) Y_{film} (0,0)
\end{equation}
is of interest, e.g., for fluid \cite{krech-1}, superfluid
\cite{garcia-chan}, and superconducting \cite{williams} films.

The universality of the scaling functions $Y$ of Eqs.
(\ref{gleichung1}) or (\ref{gleichung6}) was supposed to be valid
for all systems in a given universality class
\cite{privman-1,privman-fisher} including anisotropic lattice
systems provided that an appropriate rescaling of the lattice
spacings (or length $L$) is performed \cite{privman-fisher}. This
appears to be consistent with existing studies of finite-size
effects in anisotropic systems where it was stated that isotropy
can be restored asymptotically by an anisotropic scale
transformation
\cite{binder-etc,hucht,kim,yurishchev,cardy-1983,barber,indekeu,karevski,cardy-1987}.

We have found that this picture of finite-size effects in
anisotropic systems, though valid in special cases, is, in
general, not correct. In the present paper we show, for periodic
boundary conditions, that Eqs. (\ref{gleichung1}) -
(\ref{gleichung6}), though valid for isotropic systems and for
systems with cubic symmetry in the range $L/\xi \lesssim O (1)$
\cite{chen-dohm-2000}, are not universally valid for the
anisotropic systems of the type described above (e.g. spin models
with NN and ANNN interactions on simple-cubic lattices) although
these systems belong to the same universality class as isotropic
systems. In such anisotropic systems the finite-size scaling
functions depend, in general, on additional nonuniversal
parameters (apart from $C_1$ and $C_2$), even after a rescaling of
the lattice spacing or of the length $L$. Thus, in general,
two-scale factor universality and isotropy cannot be restored and
the notion of a universality class is only of restricted relevance
for the scaling functions of confined systems.

We shall prove our claims within the O$(n)$ symmetric $\varphi^4$
field theory with the spatially anisotropic Hamiltonian (at $h =
0$)
\begin{gather}
 H (r_0, u_0, \Lambda; {\bf A}; V; \varphi) =
\nonumber \\
\int\limits_V d^d x \left[\frac{r_0}{2} \varphi^2  +
\sum_{\alpha,
\beta}^d \frac{A_{\alpha \beta}}{2} \frac{\partial \varphi}
{\partial x_\alpha} \frac{\partial \varphi} {\partial x_\beta}  +
u_0 (\varphi^2)^2 \right]\label{gleichung3}
\end{gather}
for the $n$-component field $\varphi ({\bf x})$. The sum runs over
the components $x_\alpha$ of the spatial coordinates $\bf x$,
$\alpha=1,\ldots,d$. The $d \times d$ anisotropy matrix ${\bf
A}\equiv(A_{\alpha
\beta})$ is assumed to be real, symmetric and positive definite.
This model has a critical point at some value $r_0=r_{0 c } ({\bf
A};u_0,\Lambda)$ where $\Lambda$ is a (sharp or smooth) cutoff in
${\bf k}$ space. In addition to the three parameters $r_0, u_0$
and $\Lambda$ of the standard isotropic (${\bf A} = {\bf 1}$)
model, our model has $d (d+1) / 2$ nonuniversal parameters
contained in the matrix ${\bf A}$. Below we shall argue that the
non-diagonality of the anisotropy matrix ${\bf A}$ is a generic
case of real anisotropic systems. For simplicity we assume a cubic
volume, $V=L^d, \;0 \leq x_\alpha \leq L$, with periodic boundary
conditions.

First we prove that the model defined by Eq. (\ref{gleichung3})
belongs to the same bulk universality class as the standard
isotropic Landau-Ginzburg-Wilson model with ${\bf A} = {\bf 1}$.
The characteristic properties of the matrix $\bf A$ are described
in terms of the $d$ eigenvalues $\lambda_i>0$ and eigenvectors
${\bf e}_i$ defined by $ {\bf A e}_i=\lambda_i {\bf e}_i$. A
rotation by the orthogonal matrix $\bf U$ yields the diagonal
matrix $ {\bf U AU}^{-1}={\mbox {\boldmath$\lambda$}}$ with
diagonal elements $ \lambda_i$. After the transformation of the
spatial coordinates
\begin{equation}
\label{gleichungx} {\bf x'} = {\mbox {\boldmath$\lambda$}}^{-1/2}
{\bf U}{\bf x}
\end{equation}
and of the field
\begin{equation}
\label{gleichung9} \varphi'({\bf x}')=(\det{\bf
A})^{1/4}\varphi({\bf U}^{-1} {\mbox{\boldmath$\lambda$}}^{1/2}
{\bf x'}),
\end{equation}
\begin{equation}
\label{gleichung14}\det {\bf A}=\prod_{i=1}^d \lambda_i>0 \; ,
\end{equation}
the Hamiltonian (\ref{gleichung3}) becomes
\begin{eqnarray}
\label{gleichung10}H (r_0, u_0, \Lambda; {\bf A}; V;  \varphi) =
H' (r_0, u_0', \Lambda'; V'; \varphi') =
\end{eqnarray}
\begin{eqnarray}
\label{gleichungani} \int\limits_{V'} d^d x' \left[\frac{r_0} {2}
\varphi'({\bf x}')^2 +  \frac{1} {2} (\nabla' \varphi')^2 +
 u'_0 (\varphi'^2)^2 \right]
\end{eqnarray}
with the changed four-point coupling
\begin{equation}
\label{gleichung-14} u'_0=(\det{\bf A})^{-1/2}u_0 \;,
\end{equation}
with the changed (non-cubic) volume
\begin{equation}
\label{gleichung-13} V' = \prod_{i = 1}^d L_i' = (\det {\bf A})^{-
1/2} V \; ,
\end{equation}
\begin{equation}
\label{gleichung-15} L_i' = L \lambda_i^{- 1/2} \; ,
\end{equation}
with a transformed cutoff
$\Lambda'$ in ${\bf k}'$ space, ${\bf k}'={\mbox
{\boldmath$\lambda$}}^{1/2} {\bf U k}$, and with a critical point
at
\begin{equation}
\label{gleichung13} r_{0c}'(u_0', \Lambda')=r_{0c}({\bf
A};u_0,\Lambda) \;.
\end{equation}
The temperature variable $r_0-r_{0 c}=r_0-r_{0c}'=a_0 t$  remains
invariant under the transformation (\ref{gleichungx}) and
(\ref{gleichung9}).

According to Eq. (\ref{gleichungani}), the bulk critical behavior
of the {\it anisotropic} model $H$ with the coupling $u_0$, Eq.
(\ref{gleichung3}), can be calculated within the minimally
renormalized {\it isotropic} bulk theory $(V'\to\infty,
\Lambda'\to\infty)$ for $H'$ with the coupling $u_0'$ in $2 < d <
4$ dimensions \cite{schloms-etc}, provided that $u_0'>0$.
Specifically, the renormalized quantities of the Hamiltonian $H'$
are defined as
\begin{equation}
\label{gleichung-17-} u' = \mu^{- \varepsilon} Z_{u'}^{- 1}
Z_{\varphi '}^2 A_d u'_0 \;,
\end{equation}
\begin{equation}
\label{gleichung-19} \varphi'_R = Z_{\varphi'}^{- 1/2} \varphi'
\;,
\end{equation}
\begin{equation}
\label{gleichung-18}  r = Z_r^{- 1}(r_0-r_{0c}')
\end{equation}
with $A_d = \Gamma (3-d/2) 2^{2-d} \pi^{- d/2} (d-2)^{-1}$ and
\begin{equation}
\label{gleichung19} r_{0c}'=(u_0')^{2 / \varepsilon}
S(\varepsilon) \; ,
\end{equation}
$\varepsilon=4-d$, where $S (\varepsilon)$ and the $Z$ factors
$Z_i (u', \varepsilon)$ depend on $\varepsilon$ and $u'$ in the
same way as they depend on $\varepsilon$ and $u$ in the standard
$({\bf A}={\bf 1}, V \to \infty, \Lambda \to \infty)$ theory
\cite{schloms-etc}, with an identical fixed point value
$u'^*=u^*$. This statement applies also to the field-theoretic
functions $\zeta_r (u')$ and $\zeta_{\varphi'} (u')$ which
determine the critical exponents $\nu$ and $\eta$. This proves
that the critical behavior of $H$ and $H'$ belongs to the same
universality class in the whole range of ${\bf A}$ where $\det
{\bf A} > 0$.

Our model, Eq. (\ref{gleichung3}) with ${\bf A} \neq c_0{\bf 1}$,
can be considered as the continuum version of a $\varphi^4$
lattice Hamiltonian $H_{lattice}$ with short-range interactions
$J_{ij}$ (see, e.g., Eq. (\ref{gleichung50}) below for a lattice
model with a single lattice constant $\tilde a$). Non-cubic
anisotropies may arise either from a non-cubic lattice structure
or from non-cubic interactions on a cubic lattice (as an example
see Eq. (\ref{gleichung32}) below) or from both types of
anisotropies. In some range of ${\bf A}$ near ${\bf A} \approx c_0
{\bf 1}$ with $c_0>0$, $H_{lattice}$ and $H$ belong to the same
universality class. Note, however, that in general $r_{0c,
lattice}(J_{ij}; u_0, \tilde a) \neq r_{0c} ({\bf A}; u_0,
\Lambda)$.

In order to elucidate the effect of the non-diagonality of the
anisotropy matrix ${\bf A}$ we first discuss the bulk
order-parameter correlation function for $T \geq T_c$
\begin{equation}
\label{gleichung15} G ({\bf x};{\bf A}, u_0) \equiv <\varphi({\bf
x})\varphi({\bf 0})>_{H}
\end{equation}
where $< ... >_H$ means an average with the exponential weight
$e^{- H}$. Equations (\ref{gleichungx}), (\ref{gleichung9}) and
(\ref{gleichungani}) imply
\begin{equation}
\label{gleichung G}  G ({\bf x};{\bf A},u_0)  = (\det{\bf
A})^{-1/2} G' ({\bf x}';u'_0)
\end{equation}
where
\begin{equation}
\label{gleichung-17} G' ({\bf x}';u_0')\equiv < \varphi' ({\bf
x}') \varphi' ({\bf 0}) >_{H'} \;.
\end{equation}
The second-moment bulk correlation length $\xi'(u_0')$ associated
with $H'$ is defined by
\begin{equation}
\label{gleichung-12} {\xi'(u_0')} = \left[\frac{1} {2d} \lim_{V'
\to \infty} \frac{\int d^d x' {\bf x}'^2 G' ({\bf x}';u_0')} {\int
d^d x' G' ({\bf x}'; u_0')} \right]^{1/2} \;.
\end{equation}
For $T \geq T_c$ and $|{\bf x}'| / \xi' \lesssim O (1)$ the
asymptotic scaling form of $G' ({\bf x}'; u_0')$ reads
\cite{privman-1,privman-fisher} for $|{\bf x}'| \gg \Lambda'^{-1},
\xi' \gg \Lambda'^{-1}$
\begin{equation}
\label{gleichung corr} G' ({\bf x}'; u_0',)=A_G |{\bf
x}'|^{-d+2-\eta}\Phi (|{\bf x}'| / \xi')
\end{equation}
with a universal scaling function $\Phi$, a nonuniversal amplitude
$A_G (u_0', \Lambda')$, and with $\xi' = \xi'_0 (u_0') t^{- \nu}$,
apart from corrections to scaling. Eqs. (\ref{gleichungx}),
(\ref{gleichung G}) and (\ref{gleichung corr}) imply
asymptotically
\begin{equation}
\label{gleichung G1}  G ({\bf x};{\bf A},u_0)=A'_G
{\mid{\mbox{\boldmath$\lambda$}}^{-1/2} {\bf U} {\bf
x}\mid}^{-d+2-\eta}\Phi(\mid{\mbox{\boldmath$\lambda$}}^{-1/2}
{\bf U} {\bf x}\mid/\xi')
\end{equation}
with $A'_G=A_G(\det{\bf A})^{-1/2}$. Thus the anisotropy does not
change the universal structure of the scaling {\it function}
$\Phi$ but makes the scaling {\it argument} of $\Phi$ and the
spatial behavior of $G$ anisotropic, even right at $T_c ({\bf A})$
(see also \cite{bruce,wu,campostrini}).

Choosing $ {\bf x} = x_i {\bf e}_i$ along the principal direction
$i, i=1,...,d$ defined by the eigenvector ${\bf e}_i$, we have
$({\bf U} {\bf x})_j = x_i \delta_{ij}$ and
\begin{equation}
\label{gleichung G2}  G (x_i {\bf e}_i;{\bf A},u_0)=A'_G ({\mid
x_i \mid}/\lambda_i^{1/2})^{-d+2-\eta}\Phi(\mid
x_i\mid/\xi^{(i)}),
\end{equation}
where
\begin{equation}
\label{gleichung29} \xi^{(i)}({\bf A},u_0)=\xi_0^{(i)} t^{- \nu}
\end{equation}
are the {\it principal correlation lengths} of the anisotropic
system with the nonuniversal amplitudes
\begin{equation}
\label{gleichung30} \xi_0^{(i)} ({\bf A}, u_0)=\lambda_i^{1/2}
\xi_0' (u_0') \; .
\end{equation}
(The amplitudes $\xi'_0$ and $\xi_0^{(i)}$ may depend, in general,
also on the cutoff.) Their product
\begin{equation}
\label{gleichung-31-} V_{corr} ({\bf A})=\prod_{i = 1}^d \xi^{(i)}
\end{equation}
constitutes the appropriate measure of the correlation volume
whose shape is ellipsoidal rather than spherical. This is seen by
determining the singular part $F_s (t;{\bf A}, u_0) / V k_B T
\equiv f_s (t;{\bf A}, u_0)$ of the bulk free energy density $f =
- \lim_{V \to \infty} V^{ - 1} \ln \int {\cal D} \varphi e^{- H}$
of the anisotropic system. Using Eqs. (\ref{gleichung14}),
(\ref{gleichung10}), (\ref{gleichung-13}) and (\ref{gleichung-15})
we obtain
\begin{equation}
\label{gleichung22} f_s (t;{\bf A},u_0)=(\det{\bf A})^{-1/2}f'_s
(t;u'_0)
\end{equation}
where $f_s' (t;u_0')$ is the singular part of the bulk free energy
density $f' = - \lim_{V' \to \infty} V^{' - 1} \ln \int {\cal D}
\varphi' e^{- H'}$ associated with $H'$, Eq. (\ref{gleichungani}).
Together with Eq. (\ref{gleichung0}) for the isotropic system,
Eqs. (\ref{gleichung14}), (\ref{gleichung-13}) and
(\ref{gleichung29}) - (\ref{gleichung22}) lead to
\begin{equation}
\label{gleichung33} \lim_{t \to 0 +} f_s (t;{\bf A}, u_0) V_{corr}
({\bf A}) = Q (d, n) = {\rm universal}
\end{equation}
which is identical with Eq. (\ref{gleichung3a}).

>From Eqs. (\ref{gleichung G1}) - (\ref{gleichung30}) we see that a
complete knowledge of the asymptotic behavior of the correlation
function $G$ requires the knowledge of the $d+1$ nonuniversal
amplitudes $A'_G$ and $\xi^{(i)}_0$ and of the $d(d-1)/2$
nonuniversal parameters characterizing the directions of the $d$
eigenvectors ${\bf e}_i$. For real magnetic materials these
quantities are unknown as they depend on all microscopic details.
Furthermore, real magnetic materials may have lattice structures
and anisotropic interactions (e.g., ANN, ANNN and third ANN
interactions) corresponding to a nondiagonal matrix ${\bf A}$. It
is because of the non-diagonality of ${\bf A}$ that both a scale
transformation and a rotation is necessary and that a simple
rescaling of $d$ amplitudes is not sufficient. Clearly two-scale
factor universality is absent in the bulk correlation functions of
such anisotropic systems (e.g., metamagnets \cite {selke})
although they belong to the same universality class as isotropic
systems (e.g., fluids).

While the anisotropy does not destroy the universality of the
scaling function $\Phi$ of the {\it bulk} correlation function $G$
(in the non-exponential regime $r / \xi \lesssim O (1)$
\cite{statement}), a fundamental complication arises for {\it
confined} systems since, in general, the principal directions
${\bf e_i}$ of the intrinsic anisotropy are totally unrelated to
the orientation of the surfaces of the confining geometry (e.g.,
$L_1 \times L_2 \times ... \times L_d$ rectangular geometry). This
introduces a source of non-universality that cannot be absorbed
only by a transformation of the lengths $L_i$ of the confining
geometry or of the scaling argument. Within our model
(\ref{gleichung3}), a complete information of this source of
non-universality requires, at $h = 0$, the knowledge of $d + d
(d-1) / 2 = d (d + 1) / 2$ nonuniversal parameters (rather than
$d$ parameters). Within this model we shall show that this implies
not only the absence of two-scale factor universality but the
absence of universality itself for all finite-size scaling
functions and finite-size amplitude ratios of anisotropic systems
with non-cubic symmetry. {\it In particular, two-scale factor
universality and isotropy cannot be restored by an anisotropic
scale transformation for confined systems in rectangular
geometries with a non-diagonal anisotropy matrix ${\bf A}$}. This
is the central general result of this paper to be demonstrated in
the following on the basis of exact results in the large-$n$ limit
and of one-loop RG results for $n=1,2,3$.

First we consider the susceptibility $\chi$ (per component) of the
field-theoretic model (\ref{gleichung3}) above $T_c$ in a finite
cube with periodic boundary conditions. In the limit $n \to
\infty$ at fixed $u_0n$ it is determined by \cite{chen-1}
\begin{equation}
\label{gleichung8} \chi^{-1}  =  r_0  +  4 u_0 n  L^{-d} \sum_{\bf
k}  \left(\chi^{-1} + {\bf k \cdot A k} \right)^{-1}
\end{equation}
with ${\bf k \cdot A k}\equiv \sum^d_{\alpha, \beta} A_{\alpha
\beta} k_\alpha k_\beta$. The sum $\sum_{\bf k}$ runs over ${\bf
k}$ vectors with components $k_\alpha = 2 \pi m_\alpha / L ,
m_\alpha = 0, \pm 1,  \ldots$ up to some cutoff $\Lambda$. For $2
< d < 4 $ the asymptotic form of the correlation length $\xi'$
defined by Eq. (\ref{gleichung-12}) is $\xi'=\xi'_0 t^{1/(2-d)}$
with
\begin{equation}
\label{gleichung24} \xi_0' = (4 u'_0 n A_d a_0^{-1} /
\varepsilon)^{1 / (d-2)} \; .
\end{equation}
For large $L \gg \Lambda^{-1}$ and small $0 \leq t \ll 1$ we find
the asymptotic scaling form for $L'/\xi' \lesssim O (1)$
\begin{equation}
\label{gleichung16} \chi (t, L; {\bf A}) =  L'^{\gamma / \nu}
g_{cube} (L'/\xi'; {\bf \bar{A}}), \gamma / \nu = 2
\end{equation}
with the rescaled length
\begin{equation}
\label{gleichung12} L' = L (\det {\bf A})^{- 1/2 d}
\end{equation}
and the normalized anisotropy matrix
\begin{equation}
\label{gleichung25} {\bf \bar{A}} = {\bf A}/(\det {\bf A})^{1/d}
\end{equation}
where $g_{cube} (x; {\bf \bar A})$ is determined implicitly by
\begin{equation}
\label{gleichung17} x^{d-2}  -  g_{cube}^{(2-d)/2}  =  (4-d)
A_d^{-1} I_1 (g_{cube}^{- 1}; {\bf \bar A}) \; ,
\end{equation}
\begin{equation}
\label{gleichung18} I_j (z; {\bf \bar A}) = \int_0^\infty ds (4
\pi^2)^{- j} s^{j - 1} P (s, {\bf \bar A}) e^{- zs / 4 \pi^2},
\end{equation}
with
\begin{equation}
\label{gleichung28} P (s,{\bf \bar A})= (\pi / s)^{d/2} -
\sum_{\bf m} e^{- {\bf m \cdot \bar A m} s} \; .
\end{equation}
The sum $\sum_{\bf m}$ runs over ${\bf m} = (m_1, \ldots, m_d)$
with all integers $m_\alpha = 0, \pm 1, \ldots$. For ${\bf \bar
A}={\bf 1}, g_{cube} (x; {\bf 1}) \equiv g_{cube,iso} (x)$ is the
known scaling function of the isotropic case \cite{chen-1}. For
${\bf \bar A} \neq {\bf 1}$, however, $g_{cube} (x; {\bf \bar A})$
is nonuniversal and depends on ${\bf \bar A}$ in a highly
complicated way via the inhomogeneous $ {\bf m} \neq {\bf 0}$
modes, even after having introduced the rescaled length $L'$, Eq.
(\ref{gleichung12}). The effect of these modes depends on the
orientation of the eigenvectors ${\bf e}_i$ relative to the shape
of the confining geometry. In general this anisotropy effect
cannot be inferred from the knowledge of finite-size scaling
functions of isotropic systems of the same universality class and
cannot be described by a transformation of the argument $x$ of
$g_{cube,iso} (x)$ (unlike the case for the scaling function
$\Phi$ of the bulk correlation function $G$) or by a rescaling of
$L$.

Only in the special cases where ${\bf A} = {\mbox
{\boldmath$\lambda$}}$ is diagonal at the outset and where the
eigenvectors ${\bf e}_i$ happen to be parallel to the edges of the
confining cube, the finite-size scaling function of the {\it
anisotropic} system in a cubic geometry can be reexpressed in
terms of the scaling function of the {\it isotropic} system in a
$L_1' \times \ldots \times L_d'$ block geometry,
$L'_i=L\lambda_i^{-1/2}$. Such special cases with a diagonal
matrix ${\bf A}$ are $d = 2$ or $d = 3$ spin models on sc cubic
lattices with only NN couplings $J_x \neq J_y$
\cite{stauffer,hucht,karevski} or $J_x \neq J_y \neq J_z$
\cite{yurishchev}, respectively.

We note that a conclusive answer about the appropriate way of
rescaling the length $L$ cannot be inferred only on the basis of
the result of $\chi (0, L; {\bf A})$ at $T_c$, without further
knowledge. The same statement applies to the correlation length
$\xi_\parallel (0, L; {\bf A})$ in a $L^{d-1} \times \infty$
cylinder. It would always be possible to rewrite $\chi$ at $T_c$
in the form
\begin{equation}
\label{gleichung41} \chi (0, L; {\bf A}) = \widehat L^{\gamma/\nu}
g_{cube,iso} (0)
\end{equation}
with the amplitude $g_{cube,iso} (0)$ of the {\it isotropic}
system if all anisotropy effects are formally absorbed in the
length
\begin{equation}
\label{gleichung42} \widehat L = L' \left[g_{cube} (0; {\bf \bar
A}) / g_{cube, iso} (0) \right]^{\nu/\gamma} \; .
\end{equation}
But after the calculation of a different physical quantity at
$T_c$ it becomes obvious that this length $\widehat L$ is
inappropriate as will be demonstrated in the following.

Next we present the anisotropy effect on the finite-size scaling
function of the singular part of the reduced free energy density
per component in the large-$n$ limit for cubic geometry and
periodic boundary conditions. For $L \gg \Lambda^{-1}, 0 \leq t
\ll 1, L'/\xi' \lesssim O (1)$ we find
\begin{equation}
\label{gleichung20} f_s (t, L; {\bf A}) = L^{-d}  Y_{cube}
(L'/\xi'; {\bf \bar A}),
\end{equation}
\begin{gather}
 Y_{cube} (x; {\bf \bar A})  = -\frac {\ln 2}{2} +
\frac{(d-2) A_d} {2 d (4-d)}  \left[g (x; {\bf \bar A}) \right]^{-
d/2}
\nonumber \\
+ \frac{1} {8 \pi^2} \int\limits_0^\infty ds \left[\frac{4
\pi^2}{s} + \frac{1}{g (x;{\bf \bar A})} \right] P (s, {\bf \bar
A}) e^{- s/(4 \pi^2 g)} \label{gleichung21}
\end{gather}
where $g (x; {\bf \bar A}) \equiv g_{cube} (x; {\bf \bar A})$ is
determined implicitly by Eqs. (\ref{gleichung17}) -
(\ref{gleichung28}). The scaling function $Y_{cube} (x; {\bf \bar
A})$, including the amplitude $Y_{cube} (0;{\bf \bar A})$, is
nonuniversal. Only on the level of a lowest-mode $({\bf k}={\bf
0})$ approximation in Eq. (\ref{gleichung8}) the explicit
dependence of $Y_{cube}$ on ${\bf \bar A}$ disappears. The effect
of the ${\bf m}\neq{\bf 0}$ modes cannot be described simply by a
transformation of the scaling variable $x$ of $Y_{cube} (x; {\bf
1}) \equiv Y_{cube,iso} (x)$ of the isotropic case and it depends
on $d (d+1) / 2 - 1$ nonuniversal parameters contained in ${\bf
\bar A}$. [Equivalent parameters appear already in $G$, Eq.
(\ref{gleichung G1}).] This holds, of course, also for the
relevant case of general finite $n < \infty$ as can be shown
\cite{chen-dohm} within a one-loop RG calculation for the model
(\ref{gleichung3}). The exact scaling function $Y_{cube} (x; {\bf
\bar A})$ for general $n$ remains unknown even if the exact
scaling function $Y_{cube,iso} (x)$ were given for general $n$ and
if the exact matrix ${\bf \bar A}$ were given for a special
anisotropic system.

We note that the same rescaled length $L'$, Eq.
(\ref{gleichung12}), is employed in the scaling argument $L' /
\xi'$ of $Y_{cube}$ as in $g_{cube}$ but not in the leading
$L^{-d}$ power law of Eq. (\ref{gleichung20}). At $T = T_c$, it
would of course be possible to rewrite $f_s$ in the form
\begin{equation}
\label{gleichung45} f_s (0, L; {\bf A}) = \bar L^{-d} Y_{cube,iso}
(0)
\end{equation}
with the amplitude $Y_{cube,iso} (0)$ of the {\it isotropic}
system if the anisotropy effect is formally absorbed in the length
\begin{equation}
\label{gleichung-46} \bar L = L \left[Y_{cube} (0; {\bf \bar A}) /
Y_{cube,iso} (0) \right]^{- 1/d} \; .
\end{equation}
This length $\bar L$ differs, however, from the length $\widehat
L$, Eq. (\ref{gleichung42}), introduced formally for the
susceptibility $\chi (0, L; {\bf A})$.

As seen from our results for $\chi (t, L; {\bf A})$ and $f_s (t,
L; {\bf A})$, a possible ambiguity of defining a rescaled length
disappears after calculating the complete temperature dependence
of the finite-size scaling functions of the anisotropic system. At
the same time such results clarify whether or not isotropy can be
restored by a scale transformation. The exact analytic form of our
results (\ref{gleichung16}) and (\ref{gleichung20}) for $T \geq
T_c$ unambiguously answers this question for cubic geometry and
periodic boundary conditions. An extension of our results to
rectangular $L_1 \times L_2 \times ... \times L_d$ block geometry
\cite{chen-dohm} confirms our findings, i.e., even after a
rescaling of the lengths $L_i$ the finite-size scaling functions
remain nonuniversal for systems with a nondiagonal matrix ${\bf
A}$.

We conclude that, for rectangular geometry and periodic boundary
conditions, finite-size scaling functions are, in general, {\it
not} universally determined only by the bulk universality class
but do depend on nonuniversal parameters in a highly complicated
way if the system is anisotropic in the sense specified above. In
particular, within our model (\ref{gleichung3}), if the matrix
${\bf A}$ is nondiagonal, isotropy cannot be restored by a
rescaling of lengths \cite{recall}.

We expect that this conclusion holds also for nonperiodic boundary
conditions and for non-rectangular geometries. For example, we
expect that the universality of the amplitude $u$ of the
''corner'' term $u L^{- d} \ln L$ of the $d = 2$ and $d = 3$ free
energy density for free boundary conditions at $T_c$
\cite{cardy-etc,privman} is not generally valid for anisotropic
systems. The universality of $u$ was proven in \cite{cardy-etc}
only for {\it isotropic} $(d = 2)$ systems whereas in
\cite{privman} it was supposed to be valid "within a given RG
universality class". Furthermore, there have been calculations
\cite{cardy-1983,barber,indekeu,karevski} of edge exponents of
anisotropic spin models in wedge geometries with free boundary
conditions. It was found that the anisotropy enters explicitly
into the exponents and that it was possible to rescale lengths
anisotropically to bring the problem into an isotropic form. We
expect, however, that this is, in general, not possible for the
temperature-dependent finite-size scaling functions of lattice
systems with edges whose continuum limit yields an effective
Hamiltonian of the form of Eq. (\ref{gleichung3}) with a {\it
nondiagonal} matrix ${\bf A}$.

We briefly extend our analysis to $\infty^{d-1} \times L$ film
geometry with periodic boundary conditions, with $L$ being the
thickness in the $d^{th}$ direction. In the large-$n$ limit we
find
\begin{equation}
\label{gleichung31} f_{s,film} (t, L; {\bf A}) = L^{- d} \left[
({\bf \bar A}^{-1})_{dd})\right]^{-d /2}  Y_{film, iso} (\tilde x)
\; ,
\end{equation}
where $Y_{film, iso}$ is the scaling function for the {\it
isotropic} system, with a transformed argument $\tilde
x=\widetilde L / \xi'$,
\begin{equation}
\label{gleichung46} \widetilde L=\left[ ({\bf
 A}^{-1})_{dd})\right]^{1 /2} L
\end{equation}
and where $({\bf A}^{-1})_{dd}$ and $({\bf \bar A}^{-1})_{dd}$ are
the $d^{th}$ diagonal elements of the inverse of $\bf A$ and $\bf
\bar A$, respectively \cite{length}. In contrast with
(\ref{gleichung1}), a nonuniversal amplitude appears at bulk $T_c$
and the Casimir amplitude
\begin{equation}
\label{gleichung47} \Delta=(d - 1) \left[ ({\bf \bar
A}^{-1})_{dd}\right]^{-d /2}Y_{film, iso} (0)
\end{equation}
is nonuniversal. The simplicity of this anisotropy effect is due
to the one-loop structure of diagrams contributing to the
large-$n$ limit. From finite-size theory at order $u_0^2$
\cite{dohm} we infer a highly complicated ${\bf \bar A}$
dependence of $f_{s, film}$ for finite $n$. Furthermore we expect
that the amplitudes \cite{cardy1990} and scaling functions
\cite{eisenriegler} of density profiles in film geometry are
nonuniversal for anisotropic systems with non-cubic symmetry. More
generally, our results suggest that the feature of universality in
the theory of boundary critical phenomena \cite{diehl, dietrich,
diehl1997} as well as the notion of a "surface universality class"
and of "$(2+1)$-scale factor universality" \cite{diehl1997} need
to be reconsidered for the case of anisotropic systems.

It would also be interesting to interpret finite-size effects in
percolation problems of anisotropic systems \cite{hu,aharony} in
the light of the results of the present paper.

We illustrate our theory by the example of the Binder cumulant
ratio $U$ for $L \to \infty$ at $T_c$, Eq. (\ref{gleichung5}). We
consider a $\varphi^4$ lattice model
\begin{equation}
\label{gleichung50} H_{lattice}  =  \tilde a^d \Bigg\{\sum_i
\left[\frac{r_0}{2} \varphi_i^2 + u_0 (\varphi_i^2)^2 \right] +
\sum_{i, j} \frac{J_{ij}} {2 \tilde a^2} (\varphi_i - \varphi_j)^2
\Bigg\}
\end{equation}
with an isotropic ferromagnetic interaction $J_{ij} = J > 0$
between nearest neighbors but an anisotropic interaction $J_{ij} =
J'$ with only 6 (rather than 12) next-nearest neighbors in the
$\pm (1,1,0)$, $\pm (1,0,1)$, and $\pm (0,1,1)$ directions on a
simple-cubic lattice with a lattice constant $\tilde a$ in a cube
with periodic boundary conditions. It is expected that a
ferromagnetic critical point exists not only for $J > 0, J' \geq
0$ but also for $J > 0, J' < 0$. In the continuum limit ($\tilde
a\to 0$) this model is reduced to Eq.(\ref{gleichung3}) with
\begin{equation}
 \label{gleichung32}
 {\bf A}  = c_0 \left(\begin{array}{ccc}
  1& w & w \\
  w & 1& w \\
  w & w & 1 \\
\end{array}\right)
\end{equation}
where $w=J'/(J+2J') \leq \frac{1}{2}$ and $c_0=2(J+2J') > 0$.
(Note that the matrix $\bf A$ is diagonal for a model with
isotropic NNN interactions.) The positivity of $c_0$ requires $J'
> - J/2$. For $w\neq0$ the eigenvectors ${\bf e}_i$ are not
parallel to the cubic axes. The constant $c_0$ can be absorbed in
the bulk amplitude $\xi_0'$ of $\xi'$, (\ref{gleichung-12}), and
does not appear in
\begin{equation}
 \label{gleichung52}
 {\bf \bar A} (w) = (1 - 3 w^2 + 2 w^3)^{- 1/3} \left(\begin{array}{ccc}
  1& w & w \\
  w & 1& w \\
  w & w & 1 \\
\end{array}\right) \; .
\end{equation}
One of the eigenvalues of ${\bf A}$ vanishes at $w_c=-
\frac{1}{2}$, i.e., $J' = - J/4$ (the two other eigenvalues vanish
at $w = 1, J' = - J$). Thus $w$ may vary in the range $-
\frac{1}{2}<w\leq \frac{1}{2}$ corresponding to a line of
ferromagnetic critical points $T_c(w)$ terminating at a Lifshitz
point $T_c(w_c)$ of the $\varphi^4$ continuum model
(\ref{gleichung3}) (but not necessarily of the $\varphi^4$ lattice
model (\ref{gleichung50}) whose line of critical points $T_c (w)$
may end at a value of $w$ different from $- \frac{1}{2}$).

\begin{figure}
\includegraphics[width=85mm]{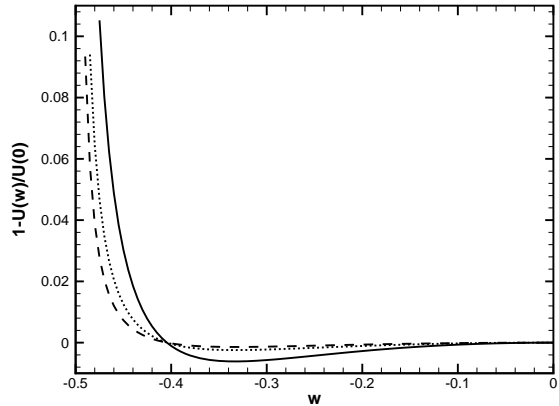}
\caption{Cumulant ratio $1-U(w)/U(0)$ vs coupling ratio
$w=J'/(J+2J')$ of the field-theoretic model, Eq.
(\ref{gleichung3}), in three dimensions for $n=1,2,3$ (solid,
dotted, dashed lines) according to Eqs. (\ref{gleichung36}) -
(\ref{gleichung55}).}
\end{figure}

>From a RG treatment of the model (\ref{gleichung3}) within the
minimal renormalization scheme in three dimensions
\cite{schloms-etc} parallel to previous work \cite{esser-etc} we
obtain $U (w)$ for $L \to \infty$ at $T_c (w)$ in one-loop order
for $n=1$ as
\begin{equation}
\label{gleichung36} U(w)= 1- \frac{1}{3} \vartheta_4 (\widetilde
Y)\left[\vartheta_2 (\widetilde Y)\right]^{-2}
\end{equation}
where
\begin{equation}
\label{gleichung38}  \vartheta_m (\widetilde Y)  =
\frac{\int\limits_{0}^{\infty} d s  s^{m} \exp
(-\frac{1}{2}\widetilde Y s^2 - s^4)} {\int\limits_{0}^{\infty} d
s \exp(-\frac{1}{2}\widetilde Y s^2 - s^4)} \; .
\end{equation}
Here the quantity $\widetilde Y$ depends on $w$ through ${\bf \bar
A} (w)$,
\begin{equation}
\label{gleichung39} \widetilde Y = - b \left\{\frac{4 \pi}{\tilde
l} [{\tilde l}^2 + I_1 ({\tilde l}^2;{\bf \bar A})] + \frac{1}{2}
+ 4 \pi \tilde l [\tilde l^4 + I_2 ({\tilde l}^2;{\bf \bar A})]
\right\},
\end{equation}
with
\begin{equation}
\label{gleichung54} b = 144 u'^* \vartheta_2 (0) \; ,
\end{equation}
\begin{equation}
\label{gleichung55} {\tilde l} = [24 \pi^{1/2} u'^{* 1/2}
\vartheta_2 (0)]^{2/3}
\end{equation}
and $u'^*=u^*=0.0412$ where $I_j (z; {\bf \bar A})$ is given by
Eq. (\ref{gleichung18}). Clearly there is no way of eliminating
the complicated {\it internal} dependence on the anisotropy matrix
${\bf \bar A}$ in Eq. (\ref{gleichung39}), thus isotropy cannot be
restored by means of a scale transformation.

We have also extended this result to general $n$ \cite{chen-dohm}.
While the $w $ dependence is weak for $- 0.4 \lesssim w \leq
\frac{1}{2}$ it becomes appreciable upon approaching $w_c = -
\frac{1}{2}$, as shown in Fig. 1 for $n=1,2,3$. This proves the
nonuniversality of $U(w)$. Similarly one can derive a $w$
dependence of the Casimir amplitude $\Delta(w)$ and of other
scaling functions. Note, however, that because of the {\it
nonuniversal character} of $U (w)$ and $\Delta (w)$, these
quantities may, in principle, differ, e.g., for the $(d = 3, n =
1)$ field-theoretic model [Eq. (\ref{gleichung3}) with the matrix
(\ref{gleichung32})], and the $(d = 3, n = 1)$ Ising model (with
NN and ANNN couplings) even if the geometries and the boundary
conditions are the same in both models.

This kind of nonuniversal finite-size effect should exist near
critical points of real systems and should be detectable in Monte
Carlo simulations of $d=2$ and $d=3$ spin models.

Support by NNSFC, by Max-Planck-Gesellschaft, by DLR and by NASA
is acknowledged.


\begin{thebibliography}{99}
\bibitem{length}
The length $\widetilde L$ is not the transformed film thickness
but rather the distance between those points on the opposite
surfaces in the transformed system that are connected via the
periodicity requirement.
\end{thebibliography}
\end{document}